# Hybrid optical-thermal antennas
# for enhanced light focusing and local temperature control

**Svetlana V. Boriskina**[*], **Lee A. Weinstein, Jonathan K. Tong, Wei-Chun Hsu, Gang Chen**

*Department of Mechanical Engineering, Massachusetts Institute of Technology,*

*Cambridge, MA 02139, USA 20036, USA*

**Abstract:**

Metal nanoantennas supporting localized surface plasmon resonances have become an indispensable tool in bio(chemical) sensing and nanoscale imaging applications. The high plasmon-enhanced electric field intensity in the visible or near-IR range that enables the above applications may also cause local heating of nanoantennas. We present a design of hybrid optical-thermal antennas that simultaneously enable intensity enhancement at the operating wavelength in the visible and nanoscale local temperature control. We demonstrate a possibility to reduce the hybrid antenna operating temperature via enhanced infrared thermal emission. We predict via rigorous numerical modeling that hybrid optical-thermal antennas that support high-quality-factor photonic-plasmonic modes enable up to two orders of magnitude enhancement of localized electric fields and of the optical power absorbed in the nanoscale metal volume. At the same time, the hybrid antenna temperature can be lowered by several hundred degrees with respect to its all-metal counterpart under continuous irradiance of $10^4$-$10^5$ W/m$^2$. The temperature reduction effect is attributed to the enhanced radiative cooling, which is mediated by the thermally-excited localized surface phonon polariton modes. We further show that temperature reduction under even higher irradiances can be achieved by a combination of enhanced radiative and convective cooling in hybrid antennas. Finally, we demonstrate how hybrid optical-thermal antennas can be used to achieve strong localized heating of nanoparticles while keeping the rest of the optical chip at low temperature.

---

[*] Correspondence and requests for materials should be addressed to S.V.B. (email: sborisk@mit.edu)

Metal nanoantennas can concentrate and locally enhance visible or near-infrared light via the excitation of localized surface plasmon resonances. The so-called 'electromagnetic hot spots' formed on the nanoantennas are widely used to localize light absorption and to enhance weak nonlinear processes such as Raman scattering, molecular fluorescence or near-infrared molecular absorption[1–6]. However, a major drawback of plasmonic nanoantennas (and plasmonic nanocircuits in general) is excessive localized heat generation in metal under light illumination[5,7–9], which may cause formation of literal hot spots – i.e., areas of high local temperature. Localized heating of plasmonic nanoantennas is very useful in some applications such as hot vapor generation[10,11], cancer treatment[12], catalysis[13–15], and nano-fabrication[16,17] and nano-manipulation[18,19]. However, in other applications such as sensing, imaging, spectroscopy, and optical signal processing the heating is typically an undesired effect that needs to be alleviated[20]. One reason for the local temperature rise of metal nanoantennas is their extremely low thermal emittance in the mid-to-far infrared frequency range. Furthermore, if the nanoantenna size is smaller than the phonon mean-free path in the material of the substrate on which is located, the conductive heat transfer from metal volume may also be reduced[21].

Prior work has already shown that metal nanoparticles generate localized temperature gradients under light illumination, which may be strong enough to create nanoscale regions of super-heated water around particles[9,22] and even to burn nanoholes in the particle substrate material[23]. Photo-induced heating of nanoantennas can also change their morphology due to the melting of metal[24]. Melting reduces the sharpness of the nanoscale features and ultimately turns nanoantennas into nanospheres if high enough temperature is reached[25]. Such morphology changes modify the antennas' spectral response and may reduce the near-field enhancement they provide[24]. Surface melting of nanoantennas can occur at temperatures much lower than the melting temperature of metal. Furthermore, for applications in sensing, spectroscopy, and near-field imaging, excessive localized heating might be harmful to the material or tissue that is being probed[26,27].

Here, we report on a new approach to *simultaneously enable intensity enhancement at the operating wavelength in the visible and adaptive control of the operating temperature* of metal nanoantennas. To achieve this goal, we proposed and designed hybrid optical-thermal nanoantennas composed of photonic (e.g., dielectric microspheres) and plasmonic (e.g., metal nanoparticles) elements that are strongly coupled both optically and via thermal conductance. The lead author and her collaborators have previously demonstrated that hybrid optoplasmonic structures enable strong spectral selectivity and dramatic near-field intensity enhancement owing to the efficient trapping and re-cycling of visible and near-infrared photons in the form of high-Q photonic modes[28–33]. Other groups have also explored field enhancement effects in hybrid antennas for applications in light focusing, sensing and emission manipulation[34–37]. Here, we show that these hybrid antennas can be designed to provide passive cooling of metal via radiative heat extraction due to enhanced thermal emittance of the polar dielectric constituent. The infrared thermal emittance can be manipulated and increased via the proper choice of both material and morphology of the dielectric element having dimensions on the scale at or below the thermal emission peak wavelength.

We also demonstrate enhanced convective cooling of hybrid antennas over their all-metal counterparts. Both, radiative and convective cooling provide additional channels to dissipate heat energy from nanoscale volumes, which become important if thermal conductance is impeded due to the nanoscale particle footprint[38] and thermal resistances in the particle support, such as thin layers, pillars, membranes and fiber tips[39–43]. Furthermore, enhanced light focusing in hybrid structures provides an opportunity to achieve the same local field enhancement at significantly lower powers of the optical pump, and thus to reduce the power absorbed in other parts of the optical chip. Enhancement of radiative and convective cooling in hybrid structures offers opportunities for the temperature reduction in optical sensors, microlasers, as well as in high-power electronic and opto-electronic chips. On the other hand, we show that at high intensity of the optical pump the enhanced light focusing in hybrid antennas can provide high enough energy input into the nanoparticle absorber to create localized super-heated spots for driving catalytic reactions.

## Results

**Absorption and light focusing efficiency of hybrid optical-thermal nanoantennas**

Strong resonant interaction of metal nanoparticles with light is mediated by the excitation of localized surface plasmon (LSP) modes. At the wavelengths of their LSP resonances, nanoparticles can exhibit scattering and absorption cross-sections significantly exceeding their geometrical cross sections[44,45]. LSP excitation also results in the strong electric field intensity enhancement in the particle near-field region. This is illustrated in Fig. 1a for the case of a plane wave illumination of a 150-nm-diameter spherical gold (Au) nanoparticle (red line). The electric field intensity enhancement in the plasmonic nanoparticle near-field is a physical mechanism underlying their applications in enhancing non-linear material response for applications such as Raman spectroscopy, infrared-absorption and fluorescence sensing. Figure 1a also shows the absorption cross-section of the Au nanoparticle as a function of wavelength (blue line), which exhibits a strong absorption peak in the visible part of the spectrum partially overlapping with the intensity enhancement peak. The geometrical cross-section of the nanoparticle is shown for comparison as the dotted blue line. The spatial distribution of the electric field intensity in the near-field of the metal nanoparticle at the maximum enhancement peak is shown in Fig. 2b, and exhibits a typical LSP mode pattern. We calculate the far-field scattering and absorption characteristics as well as the near-field intensity of plasmonic and hybrid antennas by using rigorous generalized multiple-particle Mie scattering algorithms[4] (see Methods). The dielectric permittivity of gold in the visible and infrared parts of the spectrum used in the calculations[46] is shown in Fig. S1 of supplementary information (SI).

Strong resonant absorption by nanoparticles stems from the fast decay of their LSP modes, which transfer their energy to hot charge carriers in metal. The charge carriers eventually thermalize with the metal crystal lattice, resulting in the localized particle heating[5,7–9,47]. However, in the infrared part of the frequency spectrum, the particle absorptance, and, by reciprocity, thermal emittance is severely limited by the particle size and material properties of metals (see Fig. 1c). As a result, heat dissipation from nanoparticle via thermal radiation is negligible. Figure 1d shows the equilibrium temperature of a gold nanoparticle, which would be established under illumination by a monochromatic plane wave with varying power in vacuum. The temperature is calculated by balancing the energy of the absorbed photons and the energy of thermally-emitted infrared photons (see Methods), and ignoring other energy dissipation channels. It can be seen that this temperature depends on the excitation wavelength, and reaches the melting point of gold under a relatively low-power illumination.

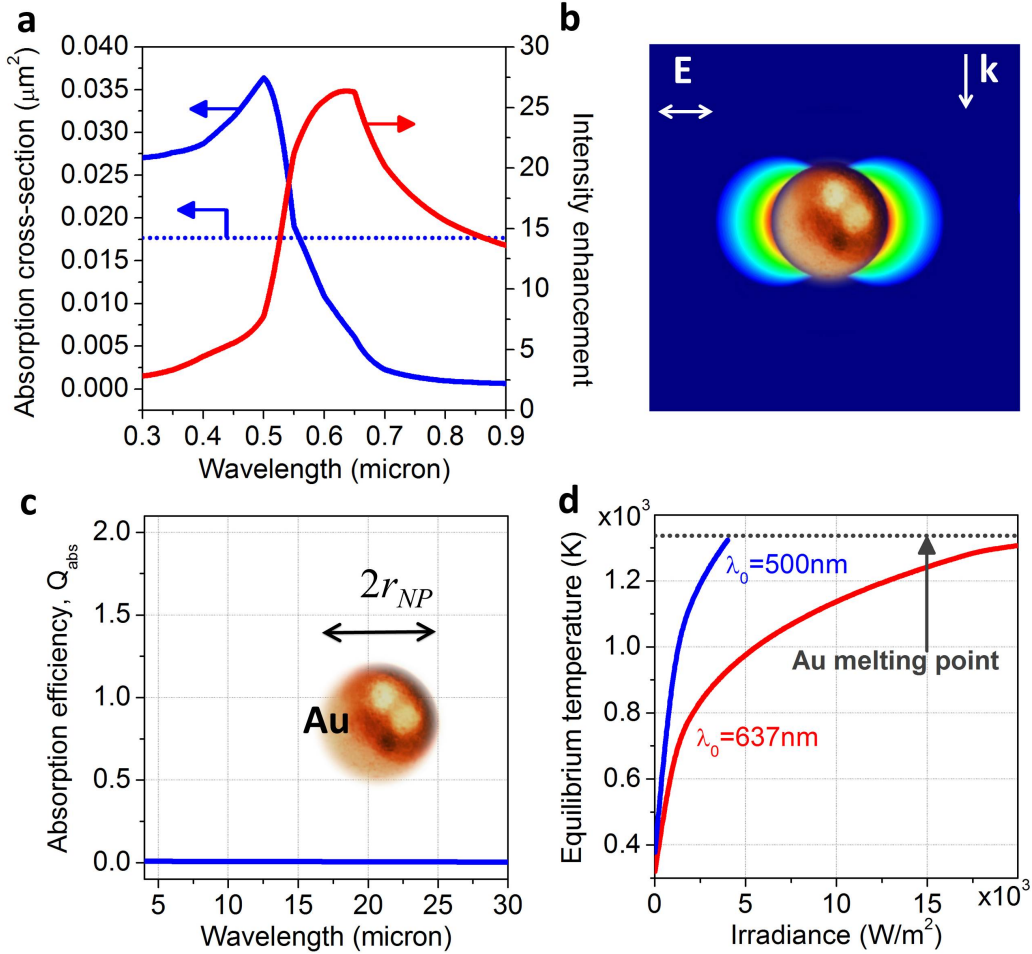

**Figure 1. Noble metal plasmonic nanoantennas amplify light intensity yet heat up by absorbing photons.** (a) Local intensity enhancement, $|E|^2/|E_0|^2$ (red line) and absorption cross-section (blue line) of an isolated 150nm-diameter Au nanoparticle in vacuum. The nanoparticle geometrical cross-section is shown for comparison (dotted blue line). (b) Electric field intensity distribution in the near field of Au nanoparticle at the wavelength of the field intensity peak in Fig. 1a (637nm). Here and in Fig. 2a the local electric field intensity is calculated inside the nanoparticle plasmonic hot-spot shown in Fig. 1b 0.5nm away from the Au surface. (c) Infrared frequency spectrum of the absorption (i.e., also emission) efficiency of the Au particle. The inset shows the particle schematic. (d) Equilibrium temperature of the particle reached under steady-state illumination by a monochromatic plane wave with varying photon flux and with a frequency centered either at the particle absorption peak of 500 nm (blue line) or at the particle intensity peak of 637nm (red line). The melting point of Au (1337K) is shown as a gray dashed line for reference.

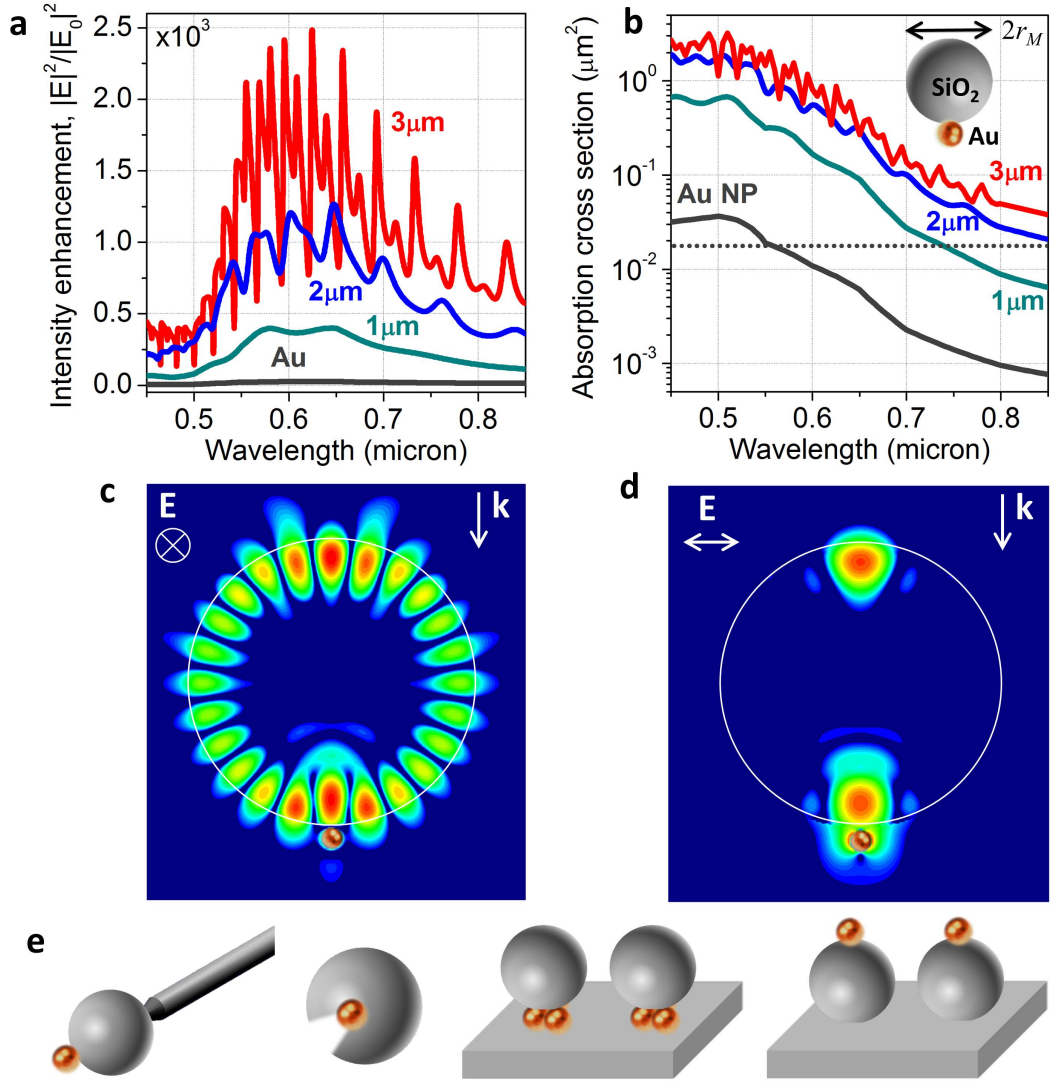

**Figure 2. Hybrid antennas enhance local field intensity and nanoparticle absorption cross-sections.** (a) Frequency spectra of the local intensity enhancement on a surface of the 150nm-diameter Au nanoparticle, which is incorporated as a part of a hybrid structure with varying diameter of the dielectric microsphere (shown as labels). The sharp intensity peaks correspond to the excitation of hybridized WG-SP modes in the hybrid antenna. The field intensity generated by a plane wave illuminating a standalone Au particle is shown as a gray line for reference. (b) Absorption cross-sections of hybrid antennas with the same parameters as in (a) as a function of wavelength. The nanoparticle geometrical cross-section is shown for comparison (dotted gray line). The inset shows a schematic of the hybrid photonic-plasmonic structure. (c,d) Electric field intensity distribution at the frequency of one of the intensity peaks in the near field of the hybrid antenna. The electric field polarization direction and the plane wave propagation direction are shown in the insets of (c,d). (e) Other possible configurations of hybrid optical-thermal antennas.

In the following, we estimate the near-field intensity enhancement and the absorption cross-section of a metal nanoparticle in a situation when it is integrated as a part of a hybrid photonic-plasmonic antenna. As the simplest hybrid antenna, we consider a dimer composed of a gold nanoparticle attached to a silica (SiO$_2$) microsphere. Frequency spectra of the intensity enhancement achievable on the surface of the metal nanoparticle integrated with a silica microsphere of varying diameter are plotted in Fig. 2a and compared to the standalone particle enhancement. The schematic of a hybrid structure is shown in the inset to Fig. 2b. Sharp peaks observed in the intensity spectrum correspond to the excitation of the whispering gallery (WG) modes in the microsphere. WG modes hybridize[48] with the localized SP modes on the Au nanoparticle, resulting in the strong resonant intensity enhancement, which can be red-shifted from the bare particle SP resonance[28–32,35]. Quality factors of WG modes grow with the increase in the microcavity radius. This translates into longer dephasing time of hybridized modes, enabling them to accumulate more energy in the near field hot spot. The near-field intensity distributions around the hybrid structure excited at the frequency of one of the resonances shown in Figs. 2c,d reveal the mechanism of the combined field enhancement and the electromagnetic hot spot formation on the nanoparticle surface. Figure 2e illustrates other possible configurations of hybrid metal-dielectric antennas, many of which can be readily fabricated by lithographic techniques or via self-assembly[49,50].

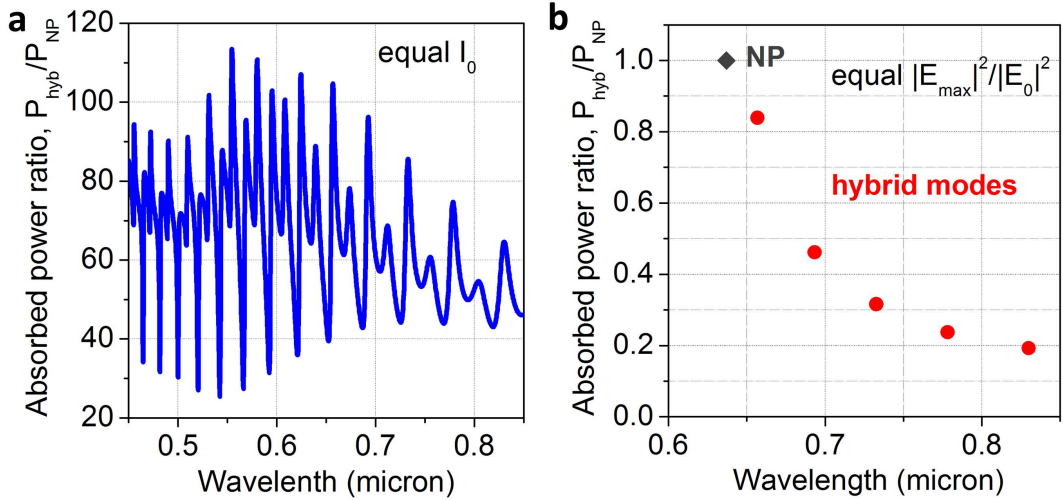

**Figure 3. Hybrid antennas increase absorbed power under constant pump irradiane yet enable absorption reduction under constant local field intensity via red-shift of the resonant wavelengths.** (a) The ratio of the optical power absorbed in the metal part of the hybrid antenna to the power absorbed in a standalone Au nanoparticle under equal irradiance as a function of wavelength. (b) The absorbed powers ratio under the condition of the equal local intensity of the electric field generated on the surface of the nanoparticle at the pump wavelength.

Figure 2b shows the frequency spectra of the absorption cross-sections of hybrid nanoantennas. It can be seen that the enhanced intensity generated in the particle near-field is accompanied by the comparable increase of its absorption cross-section. As the material dissipation losses in silica are negligible in the visible and near-infrared frequency ranges[51], all the absorption occurs in the Au nanoparticle. In effect, the glass microsphere just acts as the resonant lens that amplifies and localizes the optical signal. We verified this assumption by comparing the absorption cross-section of a hybrid structure with that of a SiO$_2$ microsphere of the same size. The results of the absorption calculations are plotted in Supplementary Fig. S2, which clearly demonstrates that the incoming light is absorbed in the gold nanoparticle volume.

To estimate the absorption enhancement, we plotted in Fig. 3a the ratio of the power absorbed in the hybrid antenna with a 3-micron silica sphere to that absorbed in a standalone Au nanoparticle. It can be seen that hybrid antennas can resonantly increase absorbed power by up to two orders of magnitude if they are illuminated by the optical pump of the same intensity as in the case of the standalone nanoparticle. However, to achieve the same local field intensity on the nanoparticle integrated into a hybrid antenna, the pump irradiance can be reduced by two orders of magnitude, which would significantly lower the power absorbed in other areas of the optical chip. The absorbed power can be further reduced by shifting the hybrid modes resonances away from the metal absorption peak while maintaining constant local field intensity. This is illustrated in Fig. 3b, which shows the reduction of the absorbed power in the 3-micron hybrid antenna operating at progressively red-shifted hybrid resonances.

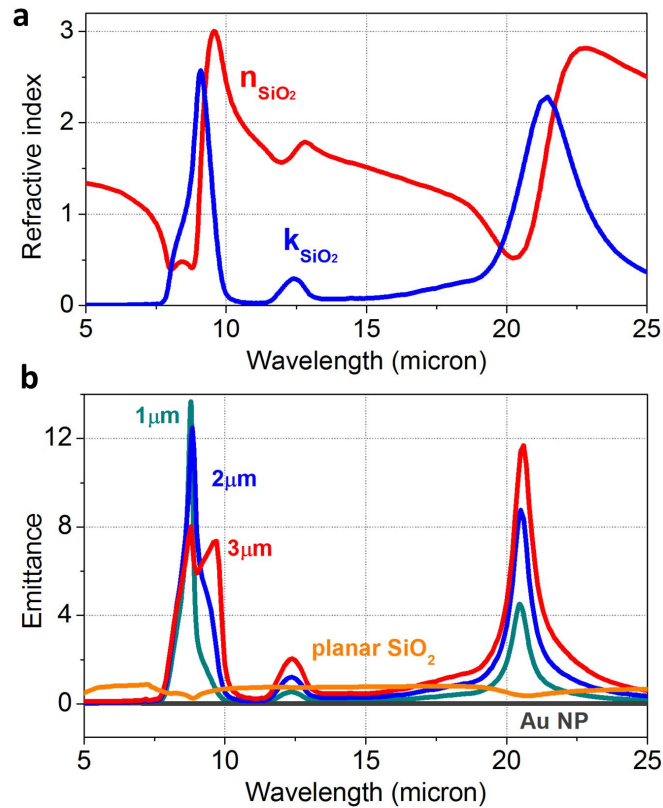

**Figure 4. Hybrid antenna provides strong resonant infrared thermal emittance.** (a) Frequency dispersion of the complex refractive index of silicon dioxide[51], which features the negative dielectric permittivity bands ($k_{SiO_2} > n_{SiO_2}$) in the infrared. (b) Frequency spectra of the infrared effective emittance of the hybrid antennas with varying diameter of the dielectric sphere (shown as labels). The emittance spectra of a $SiO_2$ surface and of a standalone Au nanoparticle are shown for comparison as orange and gray lines, respectively. For proper comparison, effective emittances normalized to the equivalent surface area are plotted for all the antenna configurations in (b).

## 2. Radiative cooling of hybrid nanoantennas

Polar dielectrics such as $SiO_2$ support surface phonon polariton modes in the infrared (IR) frequency range[52], which are akin to the surface plasmon modes supported by metals in the visible. The complex refractive index of $SiO_2$[51] is shown in Fig. 4a. It exhibits two frequency bands in the mid and far infrared frequency range, where silica behaves as metal, and is characterized by a negative dielectric permittivity. As a result, dielectric antennas can be engineered to exhibit resonant phonon-polariton-enhanced infrared absorption cross-sections. The sub-wavelength-scale polar dielectric elements can even have absorption cross-sections exceeding their geometrical cross-sections[53,54]. By Kirchhoff's law, this translates into enhanced resonant thermal emission efficiency (see Methods). Thermal emission from resonant dielectric antennas with the sizes below the dominant thermal wavelength at a given temperature is enhanced by the photon localization effects.[54–56] This favorably distinguishes resonant thermal antennas from planar radiative coolers explored before,[57–61] and makes them promising candidates to efficiently dissipate power from nanoscale high-temperature areas.

Figure 4b compares the thermal infrared emittance of the planar silica surface (orange line) with an equivalent *effective* emittance parameter of hybrid nanoantennas with varying diameter of the silica sphere (see Methods). It can be seen that at the frequencies of the localized phonon-polariton modes the effective emittance of hybrid antennas can exceed 1, which is the maximum achievable value for the emittance from a surface of a bulk blackbody. The hybrid antennas with the smallest dielectric microspheres yield the sharpest resonant peaks in the emittance spectra, while larger hybrid structures exhibit more broadband emittance enhancement (Fig. 4b). In contrast, a planar surface of bulk $SiO_2$ material exhibits suppressed emittance in the same wavelength ranges. These dips in emittance stem from the momentum mismatch between surface phonon polariton modes on a planar surface and propagating photons, which inhibits thermal radiation from the surface.

Figure 5 illustrates how much power can be dissipated by hybrid nanoantennas via thermal emission (solid lines) and compares this value to the power dissipated by emission from a standalone Au nanoparticle (gray line) and by the flat surface of bulk $SiO_2$ (dotted lines). In the latter case, the emitted power is calculated per area equal to the geometric cross-section of the corresponding hybrid antenna (matching color lines). The power dissipated by an Au particle was increased by a factor of 100 to plot it on the same scale. It can be seen that large resonant IR absorption cross-sections enable microspheres to dissipate more power than the flat surfaces with an equal footprint. The situation is reversed, however, for hybrid optical-thermal antennas with the smallest-size microspheres at very high operating temperatures, when more broadband absorptance of the flat surface enables emission of high-energy photons. The ratio of the emitted power to that of the flat emitter counterpart increases for larger microspheres, which also provide stronger near-field enhancement in the visible range (Fig. 2a).

In the absence of other channels of energy dissipation, the equilibrium temperature of the hybrid antenna can be calculated by balancing the absorbed optical power and the power dissipated via thermal emission (see Methods). Figure 5b illustrates how the equilibrium temperature of a hybrid optical-thermal antenna scales with the power of the optical pump and with the size of the silica microsphere as the power dissipation element. Predictably, the temperature rises with the increase of the optical pump irradiance; however this rise has a less steep slope for larger antennas. Although larger antennas with higher-order WG mode resonances increase the power absorbed in the nanoparticle (Fig. 2b and Fig. 3a), they provide even stronger increase in the energy dissipation via thermal radiation (Fig. 5a). For each plot in Fig. 5b, the wavelength of the optical pump was chosen to coincide with one of the sharp intensity enhancement peaks observed in Fig. 2a.

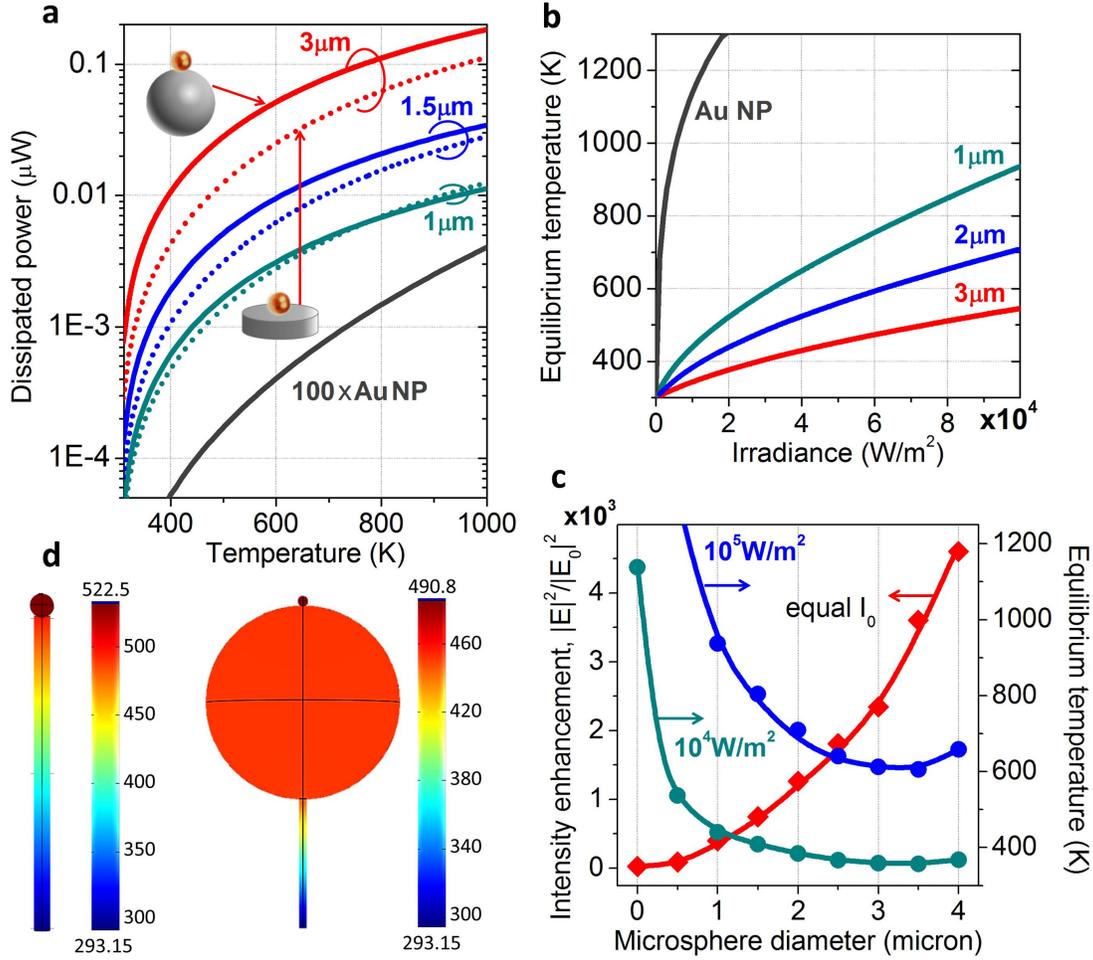

**Figure 5. Radiative cooling reduces hybrid antennas temperature by hundreds of degrees even under constant irradiance of the optical pump.** (a) The power dissipated by hybrid antennas via thermal radiation as a function of antenna temperature. Solid lines denote the power dissipated by hybrid antennas with varying radii of dielectric spheres, while dotted lines of the same color show the power dissipated by a nanoparticle on top of a planar silica surface of the same geometrical cross-section. The power dissipated by a standalone Au nanoparticle with a hundred times amplification is shown for comparison as the gray line. (b) Equilibrium temperature of the antennas reached under steady-state illumination by a monochromatic plane wave with varying photon flux. (c) Equilibrium temperature and the near-field intensity enhancement of the antennas as a function of the dielectric microsphere size. Diameter value 0 corresponds to the case of the bare nanoparticle. In (b) and (c), the optical pump has a frequency centered at each antenna's highest intensity peak (644, 642.5, 647.6, 652, 656.9, 633.6, and 641.4nm for the microsphere diameters of 1, 1.5, 2, 2.5, 3, 3.5 and 4μm, respectively). (d) COMSOL simulations of the temperature distribution in a tapered-fiber-mounted Au nanoparticle (left) and a 3-micron hybrid antenna. The support length is 2 micron and the diameter is 100nm.

Figure 5c shows how the equilibrium temperature and near-field intensity enhancement scale with the size of the thermal dielectric antenna. Here, each point on the plots (marked as a circle) is calculated by balancing the incoming power at the frequency of one of the hybridized modes of the antenna (Fig. 2a) with the total power dissipated via thermal radiation. As such, the individual points on the plot do not correspond to a smooth variation of the system parameters, and the lines connecting them are shown to highlight the trend only. Although the near-field intensity enhancement grows exponentially with the sphere size, the equilibrium temperature falls. This happens because larger microspheres dissipate more energy radiatively (Fig. 5a), while their absorption cross-sections grow at a slower pace (Fig. 2b). Thus, the hybrid optoplasmonic platforms reported in this paper offer an integrated solution to simultaneously increase the useful signal while controlling the parasitic temperature rise effect. However, with further increase of the microsphere size, absorbed power grows faster than the dissipated power, resulting in the temperature increase for larger hybrid antennas. Also, with further increase of the power of the optical pump, it becomes progressively harder to achieve simultaneous intensity increase and temperature reduction in hybrid antennas. However, as we demonstrate below, it is still possible to reduce the operating temperature of antennas via thermal emission while maintaining the same level of near-field enhancement.

The results shown in Fig. 5a-c assume that the hybrid optical-thermal antenna structure is isothermal (i.e., both the metal nanoparticle and the dielectric microsphere are at the same temperature). This implies that the optical energy absorbed by the nanoparticle is converted into heat, which in turn heats up the adjacent microsphere via conductive mechanism. Given the sizes of the antenna constituencies, this process will occur on the sub-picosecond time scale, quickly establishing the isothermal condition on the antenna and governing the thermal emission by the silica sphere. We performed COMSOL simulations to validate the isothermal structure assumption (see Methods). Previous studies have found thermal interface conductances between gold and silica to be in the range of 40-220 $MW/m^2K$[62]. The contact area between the nanoparticle and the microsphere has been estimated via the Johnson-Kendall-Roberts (JKR) model of elastic contact[63] to be 1000 $nm^2$, and we used a conservative value of the interface conductance of 40 $MW/m^2$ K. In this case, the temperature drop across the hybrid structure is less than 1 K under $10^5$ $W/m^2$ illumination (Supplementary Fig. S3). Thus the structure should be effectively isothermal, which confirms our hypothesis. Higher interface conductances can potentially be achieved by either using a thermal reflow technique, i.e., melting the gold so it makes good contact with the silica, or by applying a binding gel or grease between the particles which would act as an adhesive.

If nanoantennas are integrated on the optical chip, heat can dissipate via thermal conduction[20]. However, dissipation by conduction may heat other areas of the chip and/or molecular targets being probed by a plasmonic nanosensor. It may also lead to mechanical stresses and even cracks in the supporting material[64]. In the case of the nanoparticles mounted on tips of thin probes or nanowires, heat dissipation by conduction can be further reduced leading to higher particle temperatures[40,65]. Radiative cooling can provide an additional channel for heat dissipation, which is illustrated in Fig. 5d for an Au nanoparticle probe mounted on the tip of a tapered silica fiber. The temperature distribution has been calculated by using COMSOL Multiphysics software (see Methods). Although in this case the isothermal condition does not hold, it can be seen that replacement of the nanoparticle with a hybrid antenna probe enables reduction of operating temperature. The calculations were done assuming that both structures provide equal near-field intensity in the plasmonic hot spot. The absorbed power was 1μW for the nanoparticle ($1.38·10^8$ W/m$^2$ irradiance) and 0.833μW for the hybrid antenna case ($1.52·10^6$ W/m$^2$ irradiance).

## 3. Convective cooling of hybrid nanoantennas

Convection provides another cooling mechanism for reducing the temperature of nanoantennas. However, heat transfer to air molecules from nanoparticles with sizes comparable to the molecules mean free path (68nm at room temperature) is impeded[66] (see Methods). To illustrate this effect we calculated effective thermal conductivities of air for a 150nm nanoparticle and a 3micron microsphere under constant pressure of 1 atmosphere. The effective conductivities are plotted in Fig. 6a as a function of temperature and compared to the thermal conductivity of bulk air (dotted blue line). Significant decrease of the air thermal conductivity in the case of nanoscale particle volume can be clearly observed.

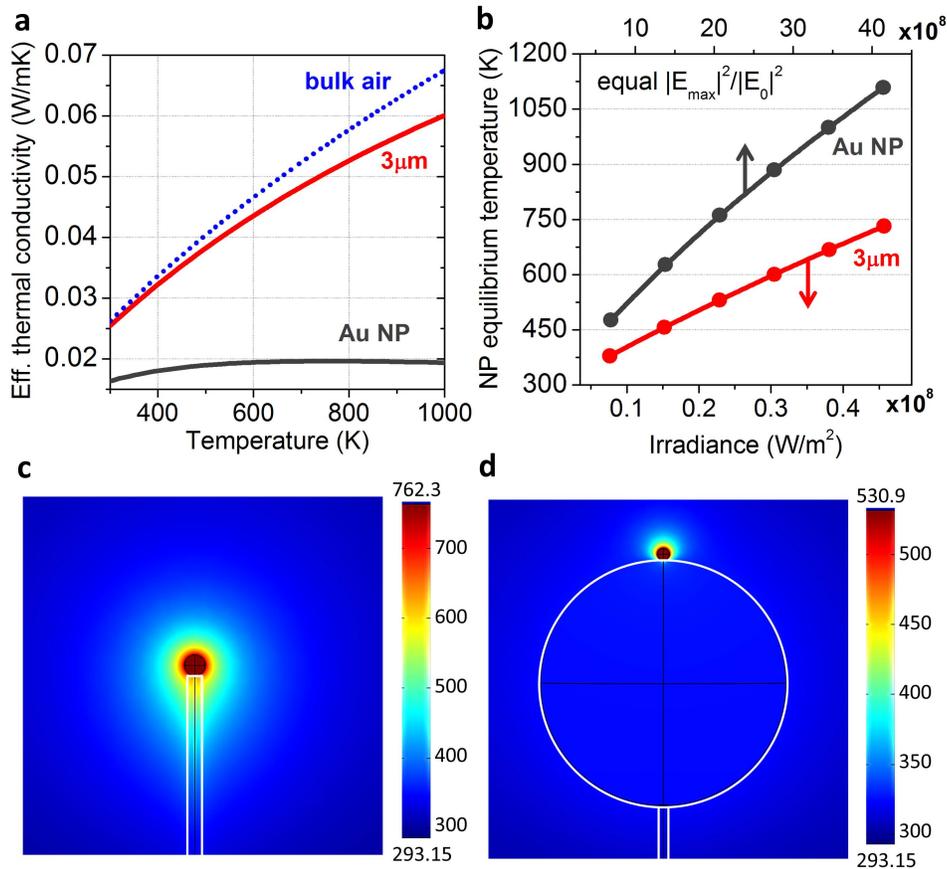

**Figure 6. Enhanced cooling of hybrid antennas via a combination of thermal radiation and air convection offers further temperature reduction.** (a) Effective thermal conductivity of air for heat removal from a 150nm nanoparticle (gray solid line) and a 3micron microsphere (red solid line) as a function of temperature. The bulk air conductivity is also shown (blue dotted line). (b) COMSOL simulations of the equilibrium temperature of the

standalone nanoparticle (gray) and the nanoparticle as a part of a hybrid antenna (red) in the same configuration as in Fig. 5d but in the presence of air convection and under the condition of equal field intensity in the plasmonic hot spot. The corresponding scales of irradiance for both cases are shown at the top and at the bottom of the plot. (c,d) Temperature distributions around the nanoparticle for the absorbed power of 15μW for the standalone nanoparticle ($20.7\cdot10^8$ W/m$^2$ irradiance) and 12.5μW for the hybrid antenna ($22.8\cdot10^6$ W/m$^2$ irradiance).

Higher effective thermal conductivity of air as well as the increase of the surface area accessible for convection in hybrid antennas enables significant reduction of their operating temperatures, while maintaining the same near-field intensity enhancement. This is illustrated in Fig. 6b, which shows how the equilibrium temperature scales with the intensity of the optical pump. Figures 6c,d offer COMSOL-calculated snapshots of the temperature distribution across the antenna volumes. Note that in this case the isothermal condition is also not applicable, and the microsphere has lower equilibrium temperature than the plasmonic nanoparticle.

## Discussion and conclusions

The described mechanisms of the signal enhancement and radiative cooling are equally applicable to more complicated optoplasmonic structures (some examples are shown in Fig. 2e), and to plasmonic particles and dielectric microcavities of various shapes and material composition. The spectral range of thermal emission can be further increased by combining several polar dielectrics (e.g., SiC and TiO$_2$) and/or metal oxides that are transparent in the visible yet exhibit plasmonic activity in the infrared (e.g. ITO).[66]

The range of hybrid structures providing both near-field intensity enhancement and thermal emission cooling mechanism also includes polar dielectric antennas deposited on top of metal films. Such a configuration has been shown to successfully launch surface plasmon polariton modes on the metal surface, as the excitation of trapped optical modes in dielectric antennas provides momentum matching to the plasmon polaritons[67]. Our COMSOL calculations (see Methods) also predict that such a structure will exhibit strong infrared thermal emission. The surface thermal emittance cannot exceed 1 in this case; however, the emittance spectra feature multiple coupling-induced resonant peaks across a broad range of infrared photon wavelengths (Supplementary Fig. 4), which should contribute to strong energy dissipation through radiative channels.

The strong light localization and enhancement achievable in hybrid antennas under lower operating temperatures is expected to benefit a wide range of applications in plasmon-enhanced spectroscopy, sensing and imaging. The same conceptual approach to the radiative cooling can also be used to design passive radiative coolers for nano- and micro-lasers. These lasers often experience significant thermal issues during their operation, and their performance can be improved by adding an extra channel for heat dissipation[68].

It should be also noted that the proposed radiative cooling platform does not need to use the sky as the cold reservoir to radiate into, which is different from previously reported radiative coolers for photovoltaic cells[69] and smart buildings[54,57]. Instead, it uses the ambient medium at room temperature as a heat sink, which sets the lower bound for the equilibrium temperature. However, if outdoor operation is possible for a specific application, the use of the sky as a heat sink can help to further reduce the operation temperature, possibly even below the ambient level[57].

Finally, we note that strong absorption enhancement in hybrid nanoantennas also offers an opportunity to generate localized areas of heat at high levels of optical pumping, when neither thermal radiation nor air conduction can compete with the absorption-generated localized heating of the nanoparticle. This situation is illustrated in Fig. 7, which compares localized temperatures that can be generated under the same level of irradiance on a 150nm Au nanoparticle on top of a glass substrate and a hybrid antenna with 3micron glass sphere on top of the same substrate. In this case, the thermal conduction to the substrate is the dominant channel of energy dissipation, which prevents the temperature rise above just a few degrees. However, as the hybrid antenna provides orders-of-magnitude enhancement of the absorbed power under the same intensity of the pump, the temperature of the nanoparticle can be raised significantly (Figs. 7a,b), providing opportunities for local enhancement of catalytic reactions[14,65,70].

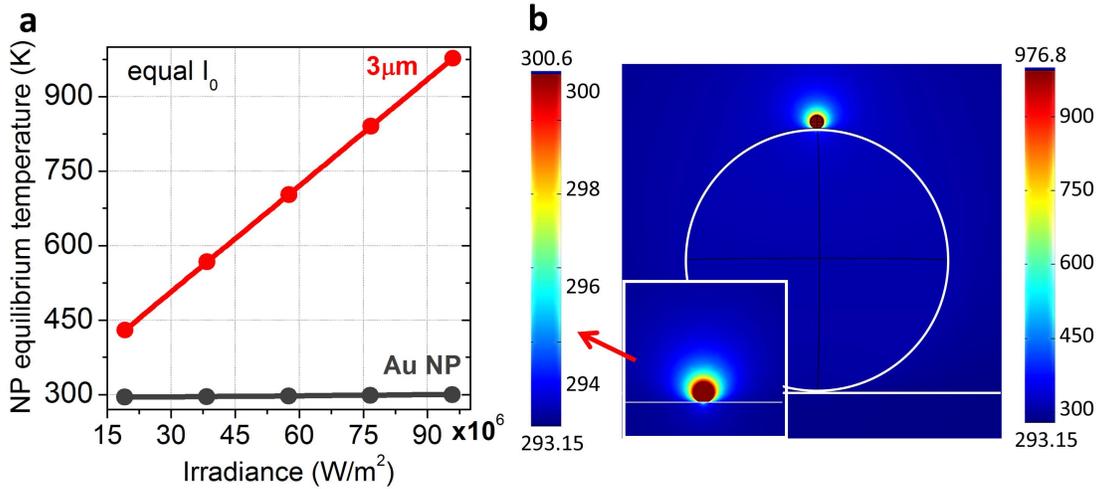

**Figure 7. Enhanced light focusing in hybrid antennas enables local temperature increase under high power optical pumping.** (a) COMSOL simulations of the equilibrium temperature of the standalone nanoparticle (gray) and the nanoparticle as a part of a hybrid antenna (red) on the silica substrate in the presence of air convection and under the condition of equal optical irradiance. (b) Temperature distributions around the nanoparticle under $9.4 \cdot 10^7$ W/m² irradiance.

## Methods

**Particle thermal emittance.** The total energy absorbed by a spherical nanoparticle of radius $r_{NP}$ irradiated by a light beam with wavelength $\lambda_i$ and power density of $W_i$ can be calculated as follows[44]:

$$W_{abs}(\lambda_i) = \pi \cdot r_{NP}^2 \cdot Q_{abs}^{NP} \cdot W_i(\lambda_i), \qquad (1)$$

where $Q_{abs}^{NP}$ is the particle absorption efficiency ($Q_{abs}^{NP} = \sigma_{abs}^{NP}/(\pi \cdot r_{NP}^2)$), which can be larger than 1 if the particle absorption cross-section $\sigma_{abs}^{NP}$ is larger than its geometrical cross-section $\pi \cdot r_{NP}^2$. Absorption cross-sections as well as the electric field intensity in the near field of plasmonic and hybrid antennas were calculated by using rigorous generalized multiple-particle Mie scattering algorithms[4] (see below).

By reciprocity[44,71], the power emitted by the nanoparticle at temperature $T$ in the form of the thermal radiation is proportional to the particle absorption efficiency, which is equal to particle emissivity[44]:

$$W_{em}^{NP}(T) = 4\pi^2 \cdot r_{NP}^2 \int_0^\infty Q_{abs}^{NP}(\lambda) \cdot N(\lambda,T) d\lambda. \qquad (2)$$

$N(\lambda,T)$ is the amount of radiant energy per unit wavelength interval and a unit solid angle, which crosses a unit area normal to the direction of propagation in a unit time, and is described by the Planck's function[72] as follows:

$$N(\lambda,T) = \frac{2\pi c^2}{\lambda^5} \cdot \left(\exp\left(\frac{hc}{\lambda k_B T}\right) - 1\right)^{-1}. \qquad (3)$$

Here, $h$ is the Planck's constant, $c$ is the vacuum speed of light, $\lambda$ is the photon wavelength, and $k_B$ is the Boltzmann constant. In direct analogy, the power emitted by the hybrid antenna scales with its absorption efficiency as:

$$W_{em}^{hyb}(T) = 4\pi^2 \cdot R_{hyb}^2 \int_0^\infty Q_{abs}^{hyb}(\lambda) \cdot N(\lambda,T) d\lambda, \qquad (4)$$

where $R_{hyb}$ is the effective cross-section of the hybrid structure ($R_{hyb} = (3V_{hyb}/4\pi)^{1/3}$, and $V_{hyb}$ is the combined volume of all the particles comprising the structure[25]). The thermal emission of the hybrid structure in the infrared spectral range originates predominantly from the surface phonon polariton modes of the microsphere (compare to the negligible nanoparticle emission shown in Fig. 1a).

In comparison, the power radiated by a flat-surface Lambertian thermal emitter from an area of the same footprint $\pi \cdot R_{hyb}^2$ as the hybrid structure effective cross-section scales with the surface emittance $e(\lambda)$, whose value cannot exceed 1 according to Kirchhoff's law:

$$W_{em}^{suf}(T) = \pi^2 \cdot R_{hyb}^2 \int_0^\infty e(\lambda) \cdot N(\lambda,T) d\lambda. \qquad (5)$$

By comparing Eqs. (4) and (5), we can define the hybrid antenna effective emittance as $e_{eff}^{hyb}(\lambda) = 4 Q_{abs}^{hyb}(\lambda)$.

The equilibrium temperature of the antenna that would be established under continuous illumination by a plane wave with the wavelength $\lambda_i$ can be found by balancing the absorbed optical power and the power dissipated via thermal emission as follows:

$$W_{abs}(\lambda_i) = W_{em}(T) - W_{em}(T_{amb}). \qquad (6)$$

Here, $W_{em}(T_{amb})$ is the power absorbed by the antenna by harvesting the ambient thermal emission. Throughout this paper, the ambient temperature was assumed to be equal to the room temperature, $T_{amb} = 300K$.

**Generalized multi-particle Mie theory.** Generalized multi-particle Mie theory (GMT) has been used for all the calculations of hybrid nanoantennas absorption cross-sections and the near-field intensity enhancement. GMT provides an exact analytical solution of Maxwell's equations for a cluster of $L$ spheres of an arbitrary spatial configuration[25]. In the frame of GMT approach, the total electromagnetic field scattered by the composite antenna structure is constructed as a superposition of partial fields scattered by each sphere. The incident, partial scattered and internal fields are expanded in the orthogonal basis of vector spherical harmonics represented in local coordinate systems associated with individual spheres. By using the translation theorem for vector spherical harmonics, the series expansion for the partial fields of the $l$-th particle is transformed into the local coordinate systems associated with other particles. A general matrix equation for the Lorenz-Mie multipole scattering coefficients is obtained by imposing the continuity conditions for the tangential components of the electric and magnetic fields on the spheres surfaces and by truncating the infinite series expansions to a maximum multipolar order $N$.

**COMSOL calculations of emittance, heat conduction and temperature in hybrid nanoantennas**. COMSOL Multiphysics, a commercial finite element analysis software package, has been used to calculate absorption cross-sections of the dielectric antenna arrays assembled on the surface of a metal thin film (Supplementary Fig. S3). The assumption that the nano-antenna structure is isothermal was also checked by using the heat transfer module in COMSOL. In this case, COMSOL was used to solve the heat equation in the structure with a fixed temperature boundary condition on the side of the silica particle opposite the gold particle and uniform heat generation in the gold particle. Numerical values for these boundary conditions were given by the rigorous GMT calculations. Bulk thermal conductivity values were used for the gold and silica particles; however the isothermal result still held even when reductions in thermal conductivity by an order of magnitude were applied (Supplementary Fig. S2).

**Calculations of effective thermal conductivity in the quasi-ballistic regime**. For nanoparticles, the effective thermal conductivity of the air is reduced from the bulk air conductivity value $K_B$, and its scaling with the size of the heated area $d$ can be described via a suppression function $S(\eta)$, $\eta = \Lambda/d$ [73]:

$$K = K_B \cdot S(\eta); \quad S(\eta) = 1 - \frac{3}{4}\eta \cdot \left(1 - 4E_s(\eta^{-1})\right) \approx \left(1 + \frac{4}{3}\eta\right)^{-1}, \qquad (7)$$

where $E_s(x)$ is the exponential integral function and $\Lambda$ is the mean free path of air molecules, which depends on gas temperature $T$, pressure $P$, and viscosity $\mu$. The mean free path of air molecules at room temperature ($\Lambda_0$) equals to 68 nm, and at any given temperature can be calculated as: $\Lambda = \Lambda_0 \cdot (\mu/\mu_0) \cdot (P/P_0) \cdot (T/T_0)^{1/2}$ [74].

## Acknowledgements

This work was supported by the U.S. Department of Energy, Office of Basic Energy Sciences, Division of Materials Science and Engineering Award No. DE-FG02-02ER45977.

## Author contributions

S.V.B. conceived the original idea, developed the Mie-theory computational methods and software, and wrote the manuscript. L.W., J.K.T. and W.-C.H. performed COMSOL calculations for nanoantennas. All the authors contributed to scientific discussions and revisions of the manuscript.

## Additional information

**Competing financial interests:** The authors declare no competing financial interests.

# Supplementary Information to

# Hybrid optical-thermal antennas
# for enhanced light focusing and local temperature control

Svetlana V. Boriskina[*], Lee A. Weinstein, Jonathan K. Tong, Wei-Chun Hsu, Gang Chen

*Department of Mechanical Engineering, Massachusetts Institute of Technology, Cambridge, MA 02139, USA 20036, USA*

**S1. Material parameters**

In all the calculations, we used experimentally measured complex refractive index data for Au[1] and SiO$_2$[2]. The refractive index data for SiO$_2$ are shown in Fig. 3a of the main manuscript text, while the data for Au are shown in Fig. S1 below.

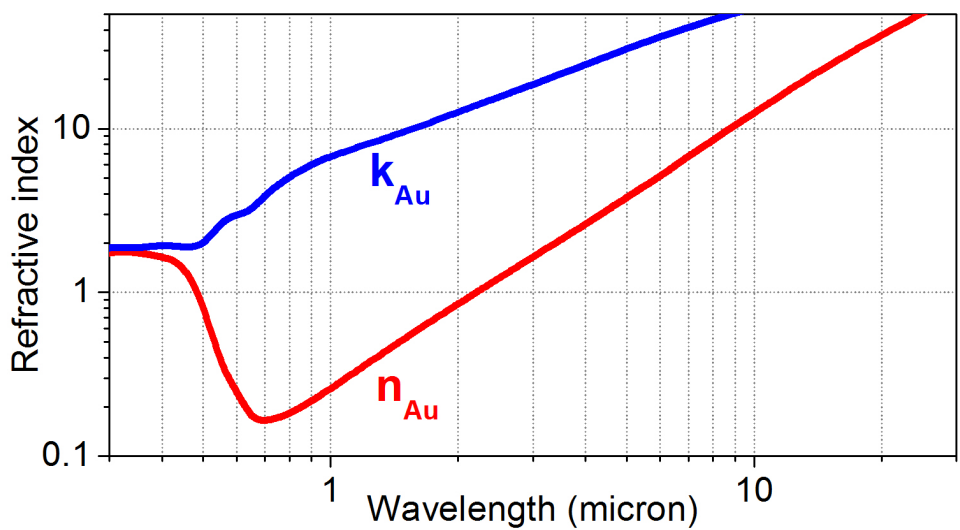

**Supplementary Figure S1.** Complex refractive index of Au in the visible and infrared spectral ranges.

[*] Correspondence and requests for materials should be addressed to S.V.B. (email: sborisk@mit.edu)

As the material dissipation losses in silica are negligible in the visible and near-infrared frequency ranges, all the absorption occurs in the Au nanoparticle. In effect, the glass microsphere just acts as the resonant lens that amplifies the optical signal. We verified this assumption by comparing the absorption cross-section of a hybrid structure with that of a bare $SiO_2$ microsphere of the same size. The results of the absorption calculations are plotted in Fig. S2, and clearly demonstrate that the incoming light is absorbed in the gold nanoparticle volume. The absorption cross-section of the bare nanoparticle is also shown for comparison.

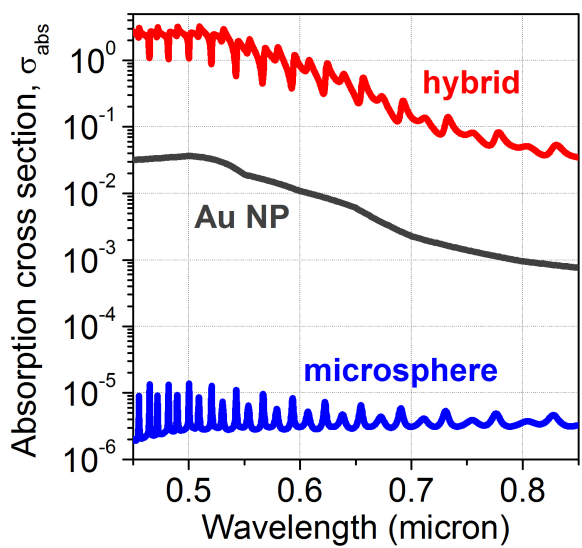

**Supplementary Figure S2.** Comparison of the absorption cross-sections of the 150nm Au nanosphere (gray line), 3-micron silica microsphere (blue line) and the corresponding hybrid antenna (red line).

**S2. Isothermal antenna condition verification**

The assumption that the hybrid antenna structure is isothermal was checked using COMSOL, a commercial finite element analysis software package. COMSOL was used to solve the heat equation in the structure with a fixed temperature boundary condition on the side of the silica particle opposite the gold particle and uniform heat generation in the gold particle. Bulk thermal conductivity values were used for the gold and silica particles, namely, 300 W/mK for gold and 1.3 W/mK for silica. We also checked that the isothermal result still held when reductions in thermal conductivity by an order of magnitude were applied to account for the possibility of the quasi-ballistic phonon transport due to the nanoscale particle footprint[3].

Figure S2 shows a screenshot of the COMSOL simulation. The figure shows a temperature heat map, with the gold particle located in the center of the screen, and the silica particle underneath. The temperature scale (in K) is the bar on the right. It can be seen that the gold particle is only about 1 degree hotter than the silica particle (so the structure can be considered effectively isothermal). The input for the thermal boundary conditions was provided through GMT calculations, which predict that the power absorbed in the Au nanoparticle under $10^5$ W/m$^2$ light illumination equals 82 nW. At the same time, the power radiatively dissipated by a hybrid system at a given temperature is obtained from the data shown in Fig. 4a. The contact area between the particle and microsphere was estimated via the Johnson-Kendall-Roberts (JKR) model of elastic contact[4] to be 1000 nm$^2$, with an interface conductance of 40MW/m$^2$K.

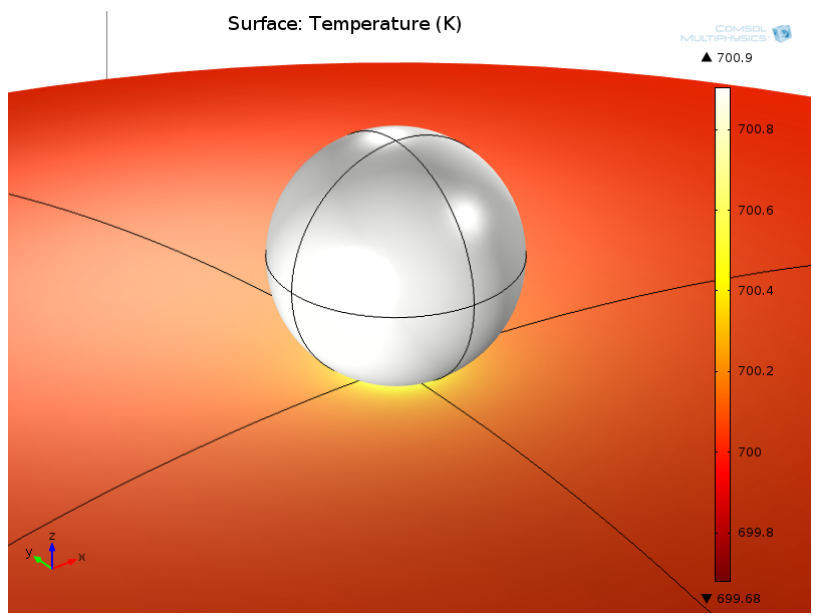

**Supplementary Figure S3.** A screenshot of the COMSOL heat conduction simulation. The 150nm diameter gold nanoparticle is at the center of the screen, and the much larger silica particle (3 micron diameter) is underneath. The temperature scale (in K) is shown on the bar on the right. The irradiance of the optical pump is assumed to be I = $10^5$W/m$^2$.

**S3. Infrared absorptance (i.e., thermal emittance) of SiO$_2$ antenna arrays on a thin Au film**

To explore other potential configurations of optical-thermal hybrid structures for light enhancement and radiative cooling, we calculated infrared surface absorptance of the dielectric antenna arrays assembled on the surface of a thin metal film (Supplementary Fig. S4). We used the wave optics module in COMSOL Multiphysics to simulate the absorptance, reflectance and transmittance for both transverse electric (TE) and transverse magnetic (TM) polarizations. To reduce numerical effort, we performed 2D calculations with periodic boundary conditions for an array of dielectric cylinders with circular cross-sections. This structure in itself can be used as a hybrid optical-thermal platform, and can also serve as a good model for an array of dielectric microspheres self-assembled on the metal surface. Such arrays have already been shown to provide enhanced coupling of propagating waves into surface plasmon polariton modes[5], and our data below demonstrate that they can also provide radiative cooling functionality. The extreme ease of fabrication that does not require any lithography steps makes these structures attractive platforms for bio(chemical) sensing applications.

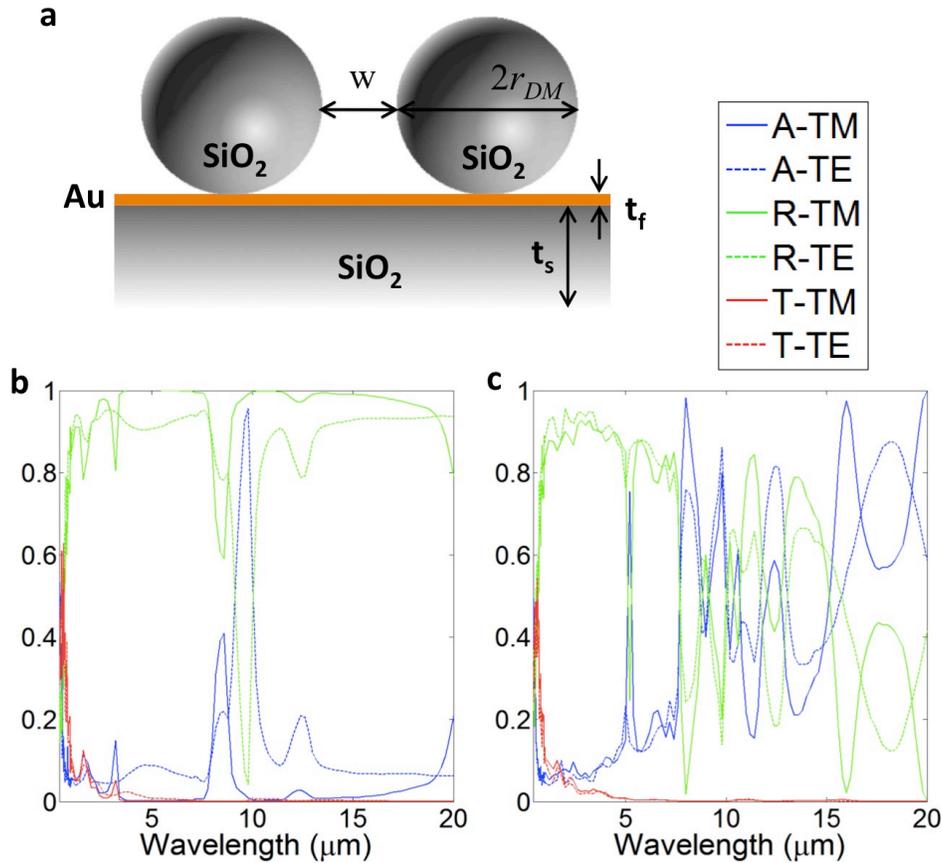

**Supplementary Figure S4.** (a) A schematic of a periodic array of silica microwires with circular cross-sections assembled on the surface of the Au thin film. (b,c) Infrared spectral characteristics of the arrays with varying diameters of the dielectric cylinders for both polarizations of light. The inter-cylinder distance is 10nm, the Au film thickness is 20nm, the SiO$_2$ substrate thickness is 2micron, and the cylinder diameters are (b) 1 and (c) 5 micron, respectively.